 \documentclass[preprint,showpacs,preprintnumbers,amsmath,amssymb]{revtex4}

\usepackage{graphicx,color}
\usepackage{dcolumn}
\usepackage{bm}

\begin{document}

\preprint{APS/123-QED}

\title{Energy-dependent Lorentz covariant parameterization\\
of the NN interaction between 50 and 200 MeV}

\author{Z. P. Li$^{1}$}
\author{G.~C. Hillhouse$^{2,1}$}
\author{J. Meng$^{1,2,3,4}$}

\affiliation{$^{1}$School of Physics, Peking University, Beijing 100871}

\affiliation{$^{2}$Department of Physics, University of Stellenbosch, Stellenbosch, South Africa}

\affiliation{$^{3}$Institute of Theoretical Physics, Chinese Academy of Science, Beijing 100080}

\affiliation{$^{4}$Center of Theoretical Nuclear Physics, National Laboratory of
Heavy Ion Accelerator, Lanzhou 730000}

\date{\today}

\begin{abstract}
For laboratory kinetic energies between 50 and 200~MeV, we focus on
generating an energy-dependent Lorentz covariant parameterization of
the on-shell nucleon-nucleon (NN) scattering amplitudes in terms of
a number of Yukawa-type meson exchanges in first-order Born
approximation. This parameterization provides a good description of
NN scattering observables in the energy range of interest, and can
also be successfully extrapolated to energies between 40 and 300~MeV.
\end{abstract}

\maketitle

\section{\label{sec:intro}Introduction}

We present an energy-dependent Lorentz covariant parameterization of
the on-shell nucleon-nucleon (NN) scattering matrix at incident
laboratory kinetic energies ranging from 50 to 200~MeV. In
particular, we focus on the relativistic Horowitz-Love-Franey (HLF)
model \cite{Ho85,Mu87,Ma96,Ma98} which parameterizes the NN
scattering amplitudes as a number of Yukawa-type meson-exchange
terms.

The original HLF model parameterized $pp$ and $pn$ scattering
amplitudes at discrete energies of 135, 200 and 400~MeV
\cite{Ho85,Mu87}. A drawback of the initial representation was that
separate fits were performed at each energy resulting in
non-systematic and unphysical trends in the various model parameters
as a function of energy. This was also undesirable in the sense that
it hindered meaningful comparisons of the off-shell properties of
the NN scattering matrix at different energies. This problem was
subsequently addressed by Maxwell who developed energy-dependent
parameterizations of the NN scattering amplitudes for two energy
ranges, namely 200 - 500~MeV \cite{Ma96} and 500 - 800~MeV
\cite{Ma98}. Since several experiments at future radioactive ion
beam facilities will be conducted at nucleon energies lower than
200~MeV, it is essential to extend the HLF model to lower energies
so as to generate reliable input for nuclear reaction models.

An attractive feature of HLF model is the existence of a
simple relationship -- lacking in conventional nonrelativistic
models -- between the Lorentz invariant NN scattering amplitudes and
mesons mediating the interaction. Moreover, this model has a physical
basis in the one-boson-exchange (OBE) picture, since the values of
the real meson-nucleon coupling constants (at energies higher than
200~MeV) are similar to those arising from more sophisticated OBE
models \cite{Ho85}. In addition to being employed as the basic NN
interaction driving the dominant reaction mechanism in several
relativistic scattering models of nucleon-induced reactions, the HLF
model has also been applied to extract relativistic microscopic
optical potentials \cite{Mu87,No05,Ma98b,Co05} for studying elastic
proton scattering from stable nuclei, as well as for generating
scattering wave functions for evaluating transition matrix elements
in various relativistic distorted wave models \cite{Ma98b,No05}.
More specifically, the latter microscopic optical potentials are
generated by folding the HLF $t$-matrix with the relevant
relativistic mean field Lorentz densities via the so-called $t
\rho$-approximation. Indeed, one of the triumphs of this folding
procedure is that the microscopic potentials, for energies from
200 to 400~MeV,  are virtually identical to the corresponding
global phenomenological Dirac optical potentials, which have
been shown to provide excellent quantitative predictions of
scattering observables for elastic proton scattering from
spin-zero stable nuclei ranging from $^{12}$C to $^{208}$Pb and
for incident energies between 20 and 1040~MeV \cite{Co93}.

One of the frontier areas of research in nuclear physics is to
understand the properties of nuclei far from the beta-stability
line, the so-called exotic or unstable nuclei. These studies provide
insight into the nuclear processes that underlie the evolution of
the stars and the origin of the elements in the cosmos. Indeed, a
number of radioactive ion beam facilities are currently under
construction to study short-lived rare isotopes. In particular,
there are plans to study exclusive proton-induced proton-knockout
reactions (using inverse kinematics) at facilities such as RIKEN and
GSI for both neutron- and proton-rich nuclei at energies lower than
200~MeV per nucleon. For the eventual interpretation of these data
it is essential to have reliable optical potentials.  Existing
global Dirac optical potentials have been constrained to reproduce
elastic proton scattering from stable nuclei \cite{Co93}, and hence
there is no reason to believe that these potentials can be reliably
extrapolated for studies of unstable nuclei. On the other hand, one
can readily extend the above-mentioned folding procedure to
calculate scattering potentials for proton-induced reactions on
exotic nuclei. Moreover, the successful application of the
relativistic $t \rho$ approximation to describe elastic proton
scattering from stable nuclei, gives one confidence in extending
this approach to study proton scattering on exotic nuclei. Two basic
ingredients underly the realization of these folding potentials,
namely a suitable analytical representation for the NN interaction
in the energy range of interest, as well as an appropriate
relativistic model of nuclear structure \cite{Me06}. Several
relativistic mean field nuclear structure models are currently being
developed for studying the structure of unstable nuclei
\cite{To03,Me06}. In this paper, however, we focus on developing a
HLF-type parameterization of the NN scattering amplitudes at
energies lower than 200~MeV. The generation of the corresponding
microscopic optical potentials will be considered in a future paper,
in which Pauli blocking corrections and nuclear medium modifications
to the NN interaction are expected to be significant and will be
taken into account~\cite{Mu87,De00,De05}.

The HLF model is described in Sec.~(\ref{sec:hlf}). The fitting
procedure and corresponding results are presented in
Secs.~(\ref{sec:fit}) and (\ref{sec:results}), respectively.

\section{\label{sec:hlf}The relativistic Horowitz-Love-Franey model}

The most general parametrization of the nonrelativistic on-shell elastic NN scattering matrix -- consistent with
rotational, parity, time-reversal and isospin invariance -- is completely determined by five complex scattering amplitudes.
Depending on the specific application of interest, the NN scattering matrix can be recast into several distinct, but
equivalent, five-term representations, one of which is the so-called Bystricky parameterization \cite{By78}, namely:
\begin{eqnarray}
  (2ik)^{-1}\hat f_{\rm{cm}}(\vec{k},\vec{k}') & =  &
  \frac{1}{2}\{(a + b)I_{1} \otimes I_{2}+(a - b)({\vec\sigma_1} \otimes I_{2}) \cdot {\hat{n}}(I_{1} \otimes {\vec\sigma_2}) \cdot {\hat{n}} + (c + d)\nonumber\\
   & \times & ({\vec\sigma_1} \otimes I_{2}) \cdot {\hat{m}}(I_{1} \otimes {\vec\sigma_2}) \cdot {\hat{m}} +  (c - d)({\vec\sigma_1} \otimes I_{2})\cdot {\hat{\ell}}(I_{1} \otimes {\vec\sigma_2})\cdot {\hat{\ell}}\nonumber\\
   & + &   e({\vec\sigma_1} \otimes I_{2} + I_{1} \otimes {\vec\sigma_2}) \cdot {\hat{n}}\}
   \label{e-bystricky}
\end{eqnarray}
where $\otimes$ denotes a
kronecker product, $\vec\sigma_{1}$ and $\vec\sigma_{2}$ refer to
the usual Pauli spin matrices associated with the projectile and
target nucleons, 1 and 2 respectively, $I_{1}$ and $I_{2}$ are the
respective $2 \times 2$ unit matrices, and the basis vectors
$\hat{\ell}$, $\hat{m}$ and $\hat{n}$ are defined as:
\begin{equation}
  \begin{array}{lll}
    {\hat{\ell}}\ =\ \frac{\vec{k}'\ +\ \vec{k}}{|\vec{k}'\ +\ \vec{k}|},
    &{\hat{m}}\ =\ \frac{\vec{k}'\ -\ \vec{k}}{|\vec{k}'\ -\ \vec{k}|},
    &{\hat{n}}\ =\ \frac{\vec{k}\ \times\ \vec{k}'}{|\vec{k}\ \times\ \vec{k}'|}
  \end{array}\;,
\end{equation}
where ${\vec{k}}$ and ${\vec{k}'}$ are the initial and final momenta
of the interacting nucleons in the NN centre-of-mass frame. The
scattering matrix is normalized such that the polarized differential
cross section for free NN scattering is given by
\begin{eqnarray}
(\frac{d\sigma}{d\Omega})_{\rm{pol}}\ =\ |\langle {\chi} ^{\dagger} _{{s'}_{1}}
{\chi}^{\dagger}_{{s'}_{2}}|\hat{f}_{\rm{cm}} | {\chi}_{{s}_{1}} {\chi}_{{s}_{2}} \rangle|^{2}
\label{e-msquare}
\end{eqnarray}
where the $\chi$'s represent the usual Pauli spinors.

For relativistic applications, the preceding nonrelativistic phenomenology, expressed
by Eq.~(\ref{e-bystricky}), can be cast into many different relativistic Lorentz invariant forms of which
the on-shell  NN matrix elements are identical to those of the nonrelativistic scattering matrix. A
convenient choice of the relativistic NN scattering matrix, commonly referred to as the IA1 representation
and originally proposed by McNeil, Ray and Wallace \cite{Mc83},
is given by:
\begin{equation}
\hat{t}_{NN} = \sum_{L=1}^{5}F_{L}(s, t, u)\lambda_L^{(1)} \cdot \lambda_L^{(2)}
\label{e-ia1}
\end{equation}
where $s$, $t$ and $u$ quantities denote the usual Mandelstam
variables, $s\equiv (k_1 +k_2)^2$, $t\equiv (k_1 - k^\prime_1)^2$,
where $u\equiv (k_1 - k^\prime_2)^2$, and $k_1\equiv(E,\vec{k})$ and
$k_2\equiv(E,-\vec{k})$ define the incident 4-momenta of the
projectile and target nucleons respectively in the center-of-mass
frame, with similar definitions holding for the outgoing (primed)
nucleons. The $L^{\rm{,s}}$ index the five Dirac matrices listed in
Table~\ref{table2_pwia}, and the dot product implies contraction of
the Lorentz indices.

\begin{table}[hbc]
\caption{Dirac matrix types parameterizing the free NN
amplitudes.}\label{table2_pwia}
\begin{tabular}{clc}
\hline\hline
\ \ \ L\ \ \    &     Lorentz type     &     Dirac matrices \\
\hline
1    &  Scalar (S)        &  $I_{4 \times 4}$\\
2    &  Vector (V)        &  $\gamma_{\mu}$\\
3    &  Pseudoscalar (P)  &  $\gamma_{5}$\\
4    &  Axial-vector (A)  &  $\gamma_{5}\gamma_{\mu}$\\
5    &  Tensor (T)        &  $\sigma_{\mu \nu}$\\
\hline
\end{tabular}
\end{table}

The transformation between the nonrelativistic on-shell NN $a$, $b$, $c$, $d$, $e$ amplitudes and
the relativistic $F_{S}$, $F_{V}$, $F_{P}$, $F_{A}$, $F_{T}$ amplitudes is readily established
by equating the matrix elements of the nonrelativistic scattering matrix, given by Eq.~(\ref{e-bystricky}),
to the matrix elements of the relativistic scattering matrix, given by Eq.~(\ref{e-ia1})
\cite{Ho85,Mu87,Ma96,Ma98}, namely:
\begin{equation}
(2i k)^{-1}\chi^\dag_{s^\prime_1}\chi^\dag_{s^\prime_2}\hat
f_{cm}\chi_{s_1}\chi_{s_2}\ =\ \bar u(k^\prime_1 , s^\prime_1)\bar
u(k^\prime_2 , s^\prime_2)\hat t_{NN}
                 u(k_1 , s_1)u(k_2 , s_2)\;,
\end{equation}
where the Dirac spinors $u$ are
normalized such that $\bar u u=1$. The corresponding $5 \times 5$ matrix ${\cal
O}_{5 \times 5}$ linking the two representations is given by
\begin{eqnarray}
\left(
\begin{array}{c}
a \\
b \\
c \\
d \\
e \\
\end{array}
\right)= i k {\cal O}_{5 \times 5} \left(
\begin{array}{c}
F_{S} \\
F_{V} \\
F_{P} \\
F_{A} \\
F_{T} \\
\end{array}
\right)
\label{e-abcdesvpat}
\end{eqnarray}
where
\begin{eqnarray}
{\cal O}_{5 \times 5}=
\left(
  \begin{array}{ccccc}
    \cos\theta  & \cos\theta & \cos\theta & -\cos\theta & -4\sin\theta\\
          1     &    -1      &      1     &       1     &      0      \\
         -1     &     1      &      1     &       1     &      0      \\
          1     &     1      &     -1     &       1     &      0      \\
    -i\sin\theta&-i\sin\theta&-i\sin\theta&  i\sin\theta&-4i\cos\theta\\
  \end{array}
  \right)
  \left(
  \begin{array}{ccccc}
    \alpha    & e\alpha+p\beta &    0       & p\alpha+e\beta & -2\beta\\
    -e\gamma  &    -\gamma     &  -p\gamma  & -\delta        &  2(p+e)\delta \\
    \alpha    & (p+e)\alpha    & 0          & (e-p)\alpha    & 2\alpha \\
    e\gamma   &  \gamma        & -p\gamma   & -\gamma        & 2\gamma   \\
    -\epsilon & -\epsilon      &     0      & \epsilon       & -2\epsilon\\
  \end{array}
  \right)\nonumber
  \end{eqnarray}
and
\begin{eqnarray}
  \alpha &=& \cos^2\frac{\theta}{2}\nonumber\\
  \beta  &=& 1+\sin^2\frac{\theta}{2}\nonumber\\
  \gamma &=& \sin^2\frac{\theta}{2}\nonumber\\
  \delta &=& 1+\cos^2\frac{\theta}{2}\nonumber\\
  \epsilon &=& \frac{E}{M} \sin\frac{\theta}{2} \cos\frac{\theta}{2}\nonumber\\
  E   &=& (k^2+M^2)^{1/2}\nonumber\\
  e      &=& \frac{E^2}{M^2}\nonumber\\
  p      &=& \frac{k^2}{M^2}\nonumber\;,
\end{eqnarray}
with $\theta$ being the NN centre-of-mass scattering angle and $M$
the free nucleon mass. The HLF  model is described in detail in
Refs.~\cite{Ho85,Mu87,Ma96,Ma98}. Here we briefly allude to the main
aspects relevant to this paper. Essentially this model parameterizes
the complex relativistic amplitudes $F_{L}(s, t, u)$ in terms of a
set of $N = 10$ meson exchanges [see Table~\ref{HLF}] in first-order
Born approximation, such that both direct and exchange NN
(tree-level) diagrams are considered separately, that is:
\begin{eqnarray}
F_{L}(s, t, u) = \frac{iM^2}{2 E k}[F_{L}^{D}(s,t) +
F_{L}^{X}(s,u)] \;, \label{e-dplusex}
\end{eqnarray}
where
\begin{eqnarray}
F_{L}^{D}(s,t) = \sum_{i=1}^{N}\delta_{L, L(i)}
\langle \vec{\tau}_{1} \cdot \vec{\tau}_{2} \rangle^{T_i}
 f^{i}(E, {|\vec{q}\;|})
\label{e-fidirect}
\end{eqnarray}
\begin{eqnarray}
F_{L}^{X}(s,u) = (-1)^{T_{NN}}\sum_{i=1}^{N}C_{L(i), L}
\langle \vec{\tau}_{1} \cdot \vec{\tau}_{2} \rangle^{T_i} f^{i}(E, {|\vec{Q}\,|})\;.
 \label{e-fierzexchange}
\end{eqnarray}
Here $T_i$ = (0,1) denotes the isospin of the $i^{\mbox{th}}$ meson,
$T_{NN}$ refers to the total isospin of the two-nucleon system,
$C_{L(i), L}$ is the Fierz matrix \cite{Ho85,Fi37}, and
\begin{eqnarray}
f^{i}({E, x}) = f^{i}_{R}({x}) - if^{i}_{I}({x})
\label{e-fiq}
\end{eqnarray}
with
\begin{eqnarray}
f^{i}_{R}({x}) = \frac{g_{i}^{2}}{{x}^{2} + m_{i}^{2}}(1 + \frac{{x}^2}{\Lambda_i^2})^{-2}
\label{e-fir}
\end{eqnarray}
\begin{eqnarray}
f^{i}_{I}({x}) = \frac{\bar{g}_{i}^{2}}{{x}^{2} + \bar{m}_{i}^{2}}(1 + \frac{{x}^2}{\bar{\Lambda}_i^2})^{-2}\;,
\label{e-fii}
\end{eqnarray}
where $x$ represents either the direct three-momentum transfer
$|\vec{q}$\;$|$ or the exchange-momentum transfer $|\vec{Q}|$,
namely
\begin{eqnarray}
\vec{q} & = & \vec{k} - \vec{k}',\nonumber\\
\vec{Q} & = & \vec{k} + \vec{k}'\;.
\end{eqnarray}
The isospin matrix elements in Eqs.~(\ref{e-fidirect}) and (\ref{e-fierzexchange}) yield
\begin{eqnarray}
\langle \vec{\tau}_{1} \cdot \vec{\tau}_{2} \rangle^{T_i} &=&
\left\{
\begin{array}{c}
1\ \ \mbox{for the exchange of}\ \ T_{i} = 0\ \  \mbox{(isoscalar) mesons}\\
1\ \ \mbox{for the exchange of}\ \ T_{i} = 1\ \  \mbox{(isovector) mesons}\\
\end{array}
\right\}\
\label{e-teq1amps}
\end{eqnarray}
for the $T_{NN} = 1$ amplitudes and
\begin{eqnarray}
\langle \vec{\tau}_{1} \cdot \vec{\tau}_{2} \rangle^{T_i} &=&
\left\{
\begin{array}{c}
1\ \ \mbox{for the exchange of}\ \ T_{i} = 0\ \  \mbox{(isoscalar) mesons}\\
-3\ \ \mbox{for the exchange of}\ \  T_{i} = 1\ \  \mbox{(isovector) mesons}\\
\end{array}
\right\}\
\label{e-teq0amps}
\end{eqnarray}
for the $T_{NN} = 0$ amplitudes.

Note that the mesons in Table~\ref{HLF} represent different Lorentz
types (S, V, P, A, T) with an isospin of either 0 or 1. The coupling
constants are complex with real and imaginary parts, $g_{i}^{2}$ and
$\bar{g}_{i}^{2}$, respectively. The imaginary couplings represent a
purely phenomenological means of parameterizing the imaginary
amplitudes. The meson propagators are
\begin{eqnarray}
\frac{1}{{x}^2 + m_{i}^2}\;, \label{e-propagator}
\end{eqnarray}
and the following monopole form factors are assumed for the meson-nucleon vertices,
\begin{eqnarray}
\frac{1}{1 + \frac{{x}^2}{\Lambda_{i}^2}}\;, \label{e-formfactor}
\end{eqnarray}
with {\it separate} masses ($m_{i}$, $\bar{m}_{i}$) and cutoff
parameters ($\Lambda_{i}$, $\bar{\Lambda}_{i}$) for the real and
imaginary parts of the amplitudes, respectively.

\section{\label{sec:fit}Fitting procedure}

Using some initial guess (see later) for the HLF model parameters,
the relativistic $F_{S}$, $F_{V}$, $F_{P}$, $F_{A}$, $F_{T}$
amplitudes are determined [from Eq.~(\ref{e-dplusex})], and these in
turn are converted to the nonrelativistic $a$, $b$, $c$, $d$ and $e$
amplitudes-- via Eq.~(\ref{e-abcdesvpat}) -- and subsequently
compared to the SP05 (Spring 2005) empirical amplitudes to establish
the goodness of the fit. More specifically, the HLF parameters are
varied to simultaneously fit the amplitudes $a$, $b$, $c$, $d$, $e$
for both $T_{NN} = (0, 1)$ at laboratory kinetic energies between 50
and 200~MeV (in 25~MeV steps) and centre-of-mass scattering angles
between 5 and 175 degrees (in 5 degree steps). Empirical values of
the $a$, $b$, $c$, $d$ and $e$ amplitudes were extracted via the
on-line Scattering Analysis Interactive Dial-in (SAID) facility. On
a more technical note, the NN amplitudes were obtained via ssh call
to the SAID facility, gwdac.phys.gwu.edu, with user id {\it said}
(no password), in which there are 4 choices for different types of
NN scattering, namely $pp$, $np$, $np1$, $np0$. For this paper we
selected the $np1$ and $np0$ options for generating the $T_{NN}=1$
and $T_{NN}=0$ amplitudes, EXCLUDING Coulomb corrections,
respectively. After choosing the isospin of interest, one is
confronted with 6 different choices for different nonrelativistic
amplitude types: 1 (VPI-H), 2 (Wolfenstein), 3 (Bystricky), 4
(Helicity), 5 (Transversity), or 6 (Transverse planar). We chose
option 2 (Wolfenstein) which actually represents the so-called
Hoshizaki amplitudes \cite{By78}. The latter $a_H, m_H, g_H, h_H$
and $c_H$ Hoshizaki amplitudes are subsequently converted to the
required Bystricky $a$, $b$, $c$, $d$ and $e$ amplitudes [in
Eq.~(\ref{e-bystricky})] using the transformation \cite{By78}:
\begin{equation}
 a=a_H+m_H,\ \ \ \ b=a_H-m_H,\ \ \ \ c=2g_H,\ \ \ \ d=-2h_H,\ \ \ \ e=2c_H\;.
\end{equation}
Separate fits were performed to the real and imaginary parts of the
amplitudes. In order for the 5 empirical amplitudes to be weighted
equally in the fit, we minimized the value of $\chi^{2}$ defined by
\begin{equation}
  \chi^2=\sum\limits_{\rm{data}}\frac{(x_{\rm{empirical}} - x_{\rm{fit}})^2}{\langle x^2_{\rm{empirical}}\rangle}
\end{equation}
employing a Levenberg-Marquadt method.
Here ${\langle x^2_{\rm{empirical}}\rangle}$ represents an
angle-averaged value. For a specific amplitude, say $a$, and $T_{\rm{lab}}$
the angle-averaged value is determined by:
\begin{equation}
  \langle
  a^2_{\rm{empirical}}\rangle=\frac{\sum\limits_{\theta=5^\circ}^{175^\circ}a^2_{\rm{empirical}}(\theta)}{N_{\rm ang}}
\end{equation}
where $N_{\rm{ang}} = 35$  represents the number of angles fitted per energy. The total number of the data points
per real or imaginary fit is 2450: 7 energies $\times$ 35 angles  $\times$ 5 amplitudes $\times$ 2 isospins.

\begin{figure}
\includegraphics[scale=1]{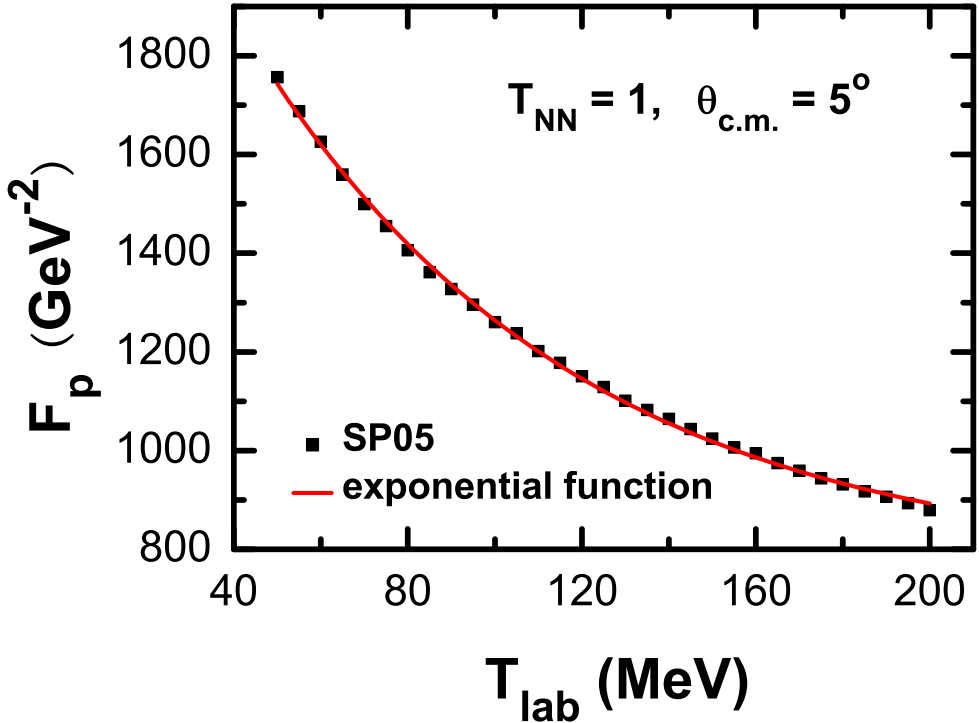} 
\caption{\label{fig:Fp-TNN1-Acm=5}(Color online) The experimental
SP05 pseudoscalar amplitudes $T_{NN}=1$ $F_p$, represented by the
solid squares, as a function of the laboratory kinetic energy
$T_{\rm{lab}}$ for a small centre-of-mass scattering angle of 5
degrees. The solid curve is an exponential function of the form
given by Eq.~(\ref{e-expfunc}).}
\end{figure}

We now discuss the general considerations for extracting the optimal HLF coupling strengths, meson masses
and cutoff parameters. The energy dependence of the coupling constants was established as follows.
The dominant contribution to the pseudoscalar $F_{P}$ amplitude is determined almost
completely by pion exchange to the direct term \cite{Ho85}, and thus according
to Eqs.~(\ref{e-dplusex}) -- (\ref{e-fii}) the amplitude $F_{P}$ at small centre-of-mass
scattering angles, and hence small values of $|\vec{q}\,|$, is approximately proportional to
$g^2_\pi$. Consequently the relationship between $F_p$ and $T_{\rm{lab}}$
(the laboratory kinetic energy) is essentially the same as the functional dependence of $g^2_\pi$ on
$T_{\rm{lab}}$. In Fig.~\ref{fig:Fp-TNN1-Acm=5}, we plot the $F_{P}$ amplitude (for $T_{NN}=1$) as
a function of $T_{\rm{lab}}$ for a small centre-of-mass scattering angle of 5 degrees.
A clear exponential energy-dependence is observed. Guided by the latter trend, we chose
the following exponential energy-dependence for all (both real and imaginary)
coupling constants, namely
 \begin{equation}
  g^{2}(E)\ =\ g^{2}_{0}[1\ +\ a_g(e^{a_{T} T_{\rm{rel}}}\ -\ 1)]
  \label{e-expfunc}
\end{equation}
where
\begin{equation}
  T_{\rm{rel}}\equiv \frac{T_0-T_{\rm{lab}}}{T_0}
\end{equation}
with $T_{0} = 200$~MeV, $T_{\rm{rel}}$ is positive in the
50 to 200~MeV energy range of interest, and $g^{2}_{0}$, $a_{g}$ and
$a_{T}$ are dimensionless parameters extracted by fitting to the
relevant data. For the real meson masses we chose the experimentally
measured values \cite{Ha02} for the $\pi$, $\omega$ and $\rho$
mesons, namely 138, 782, and 770~MeV, respectively. For the
other real and imaginary meson masses, as well as the real
and imaginary cutoff energies $\Lambda_{i}$ and $\bar{\Lambda}_{i}$, we
chose the starting the values to be the same as those published
in Table~I of Ref.~\cite{Ma96}.

Firstly, the  meson masses, coupling constants and cutoff parameters were varied
slightly to obtain the best fit to the 200~MeV data alone. Similar to
Ref.~\cite{Ma96}, the values of $\Lambda_{i}$ and $\bar{\Lambda}_{i}$ were restricted
to always exceed the respective meson masses. Thereafter, the meson masses and
cutoff parameters were kept fixed, and $a_g$ and $a_T$, which determine the coupling
constants via Eq.~(\ref{e-expfunc}), were varied to obtain the best fit to the total
data set between 50 and 200~MeV. Then all parameters (with the exception
of the meson masses) were varied to fit the total data set.

\section{\label{sec:results}Results}

The fits to the real and imaginary amplitudes yield minimum $\chi^2$
values of 10.203 and 10.798, respectively: the relative $\chi^2$
values can be obtained by dividing the latter values by 2450. The
fit to the real part of amplitude
$a$ for $I_{NN}=0$ was found to be inferior compared to
the real parts of the other amplitudes. Therefore, for
the real part, the weight of amplitude $a$ was multiplied by 200,
and the new $\chi^2$ value is equal to 16.657. The imaginary parts of
amplitudes $b$, $c$ for $I_{NN}=1$ and $a$, $b$, $c$ for $I_{NN}=0$
showed systematic deviations in the low energy range. For the imaginary
amplitudes, the parameter $a_T$ of the isovector tensor meson was
changed slightly, such that the systematic deviation nearly
disappeared, and the fits to the other amplitudes remained
satisfactory. Fitted values for the various HLF parameters are
presented in Table~\ref{HLF}.

\begin{table}
\caption{\label{HLF}Real and imaginary HLF parameters. The masses $m$ and cutoff parameters $\Lambda$
are in MeV, whereas the other parameters are dimensionless.}
\begin{ruledtabular}
\begin{tabular}{lllccccc}
\multicolumn{8}{c}{Real parameters}\\
\hline
Meson & Isospin &  Coupling type & m &$g^2_0$
 &$a_g$& $a_T$ &$\Lambda$\\
$\sigma$ & 0   & Scalar (S)  &     600  &    -8.379  &    $-5.581\times 10^{-1}$   &    $2.364\times 10^{-1}$  &   965.23\\
$\omega$ &  0  & Vector (V) &     782  &    $1.014\times 10^1$  &    $-2.219\times 10^{-2}$   &    3.793  &  1158.74\\
$t_{0}$ & 0  &  Tensor (T) &   550  &     $2.783\times 10^{-1}$  &     1.331   &    3.074  &  1955.59\\
$a_{0}$  & 0  & Axial vector (A) & 500  & $4.842\times 10^{-1}$  &     1.440   &    2.847  &  1577.53\\
$\eta$ & 0 & Pseudoscalar (P) & 950  &    $1.089\times 10^1$  &     1.553   &    1.956  &   980.99\\
$\delta$  & 1  & Scalar (S) &      500  &   $6.233\times 10^{-3}$  &     5.675   &    3.521  &  3000.00\\
$\rho$  & 1 & Vector (V) &      770  &    $-1.535\times 10^{-1}$  &     3.429   &    $1.825\times 10^{-1}$  &  3000.00\\
$t_{1}$ & 1 & Tensor (T) &      600  &    $-2.497\times 10^{-1}$  &    $5.508\times 10^{-1}$   &    3.959  &  1290.72\\
$a_{1}$ & 1 & Axial vector (A) & 650  &   -1.355  &     $2.480\times 10^{-1}$   &    3.623  &   745.19\\
$\pi$  &  1 & Pseudoscalar (P) & 138  &    $1.195\times 10^1$  &    $-1.671\times 10^{-1}$   &    $3.216\times 10^{-1}$  &   678.44\\
          \\
          \hline
          \hline
\multicolumn{8}{c}{Imaginary parameters}\\
\hline
Meson & Isospin &  Coupling type & m &$\bar{g}^2_0$
 &$\bar{a}_g$& $\bar{a}_T$ &$\bar{\Lambda}$\\
$\sigma$ & 0   & Scalar (S)  & 600  &    -2.866  &   $7.722\times 10^{-1}$   &    1.503  &   772.74\\
$\omega$ &  0  & Vector (V) & 700  &     4.415  &    $7.094\times 10^{-1}$   &    1.528  &   701.00\\
$t_{0}$ & 0  &  Tensor (T) & 750  &    $-9.114\times 10^{-1}$  &     1.217   &    1.674  &  1395.12\\
$a_{0}$  & 0  & Axial vector (A) & 750  &    -2.124  &     1.187   &    1.524  &  1364.90\\
$\eta$ & 0 & Pseudoscalar (P) & 1000 &    $1.411\times 10^{1}$  &     $3.021\times 10^{-2}$   &    4.970  &  3000.00\\
$\delta$  & 1  & Scalar (S) & 650  &     3.089  &     $9.873\times 10^{-1}$   &    $8.292\times 10^{-1}$  &   771.14\\
$\rho$  & 1 & Vector (V) &  600  &    -2.464  &      $7.620\times 10^{-1}$   &    1.176  &   795.81\\
$t_{1}$ & 1 & Tensor (T) &  750  &     $6.447\times 10^{-1}$  &     $3.316\times 10^{-1}$   &    2.977  &  1741.21\\
$a_{1}$ & 1 & Axial vector (A) & 1000 &     2.328  &     $3.885\times 10^{-1}$   &    2.915  &  1256.95\\
$\pi$  &  1 & Pseudoscalar (P) &  500  &    -5.788  &    $7.635\times 10^{-1}$   &    1.623  &  1391.26\\
\end{tabular}
\end{ruledtabular}
\end{table}

\begin{figure}
\includegraphics[scale=1.2]{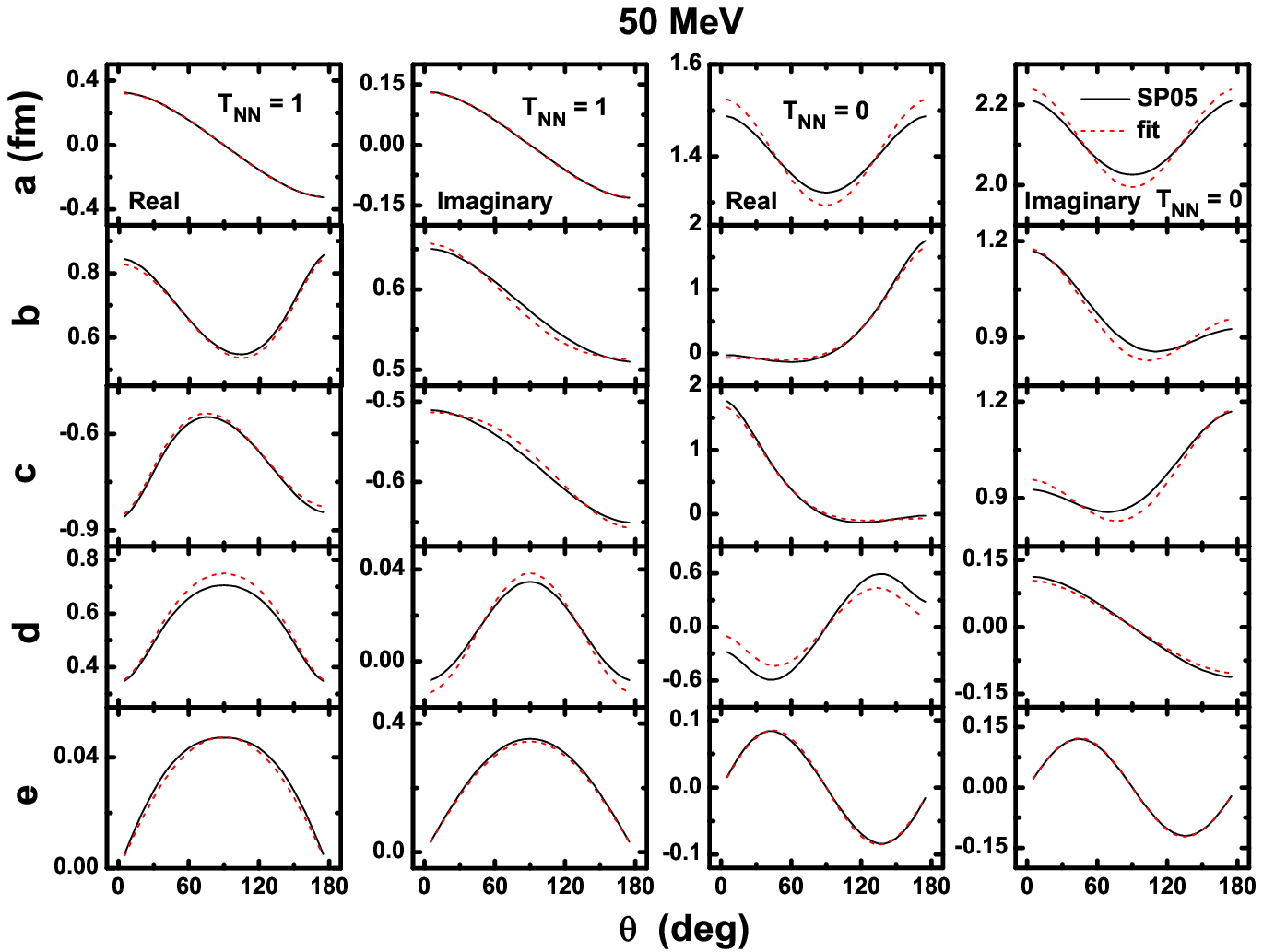} 
\caption{\label{fig:ampl-50}(Color online) Comparison of the fitted
real and imaginary ($a,b,c,d,e$) $T_{NN} = 1$ and $T_{NN} = 0$
amplitudes (dashed curve), in units of fm, with the empirical SP05
amplitudes (solid curve) \cite{Arndt} as a function of the NN
centre-of-mass angle $\theta$, in degrees, at a laboratory kinetic
energy 50~MeV.}
\end{figure}

\begin{figure}
\includegraphics[scale=1.2]{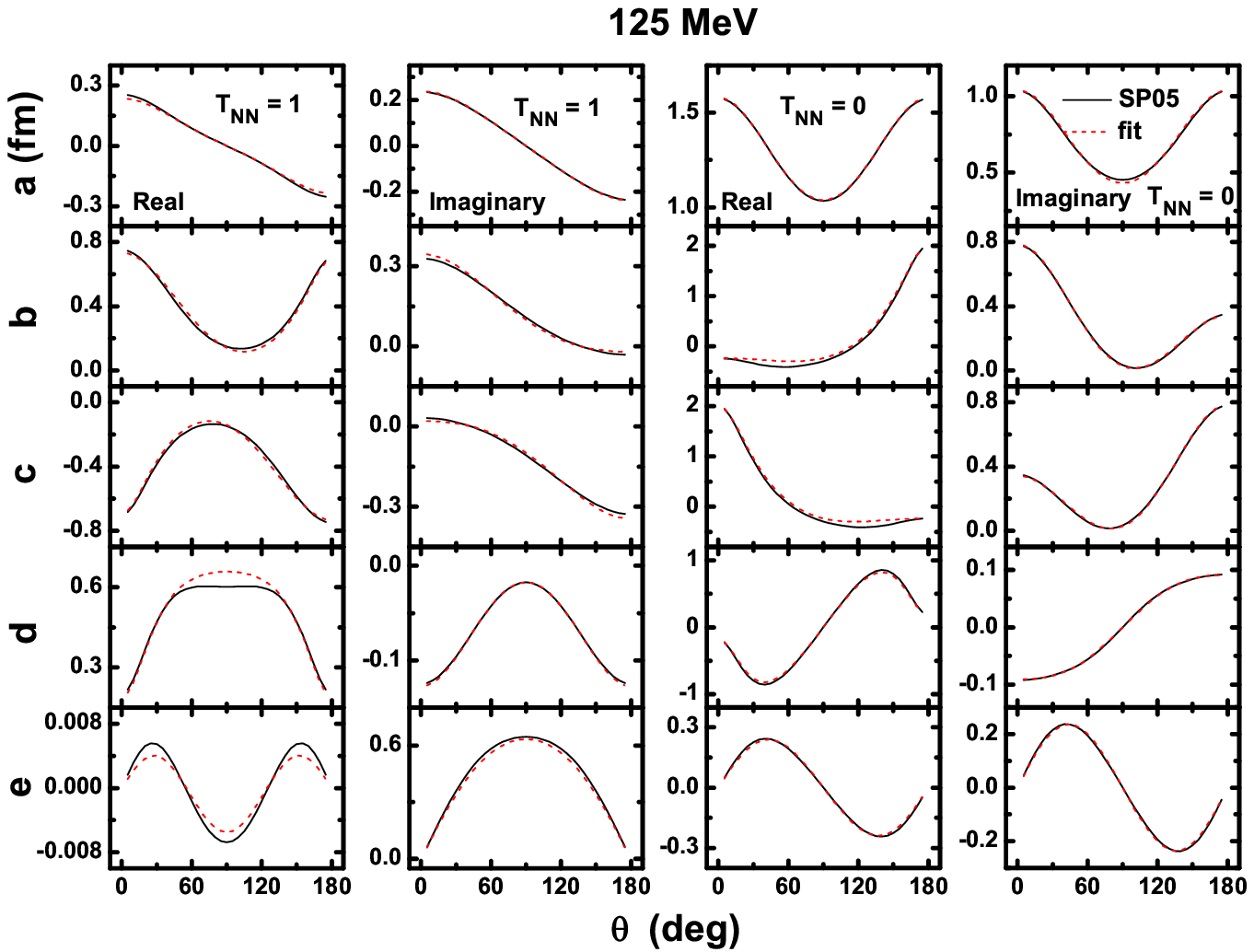} 
\caption{\label{fig:ampl-125}(Color online) Comparison of the fitted
real and imaginary ($a,b,c,d,e$) $T_{NN} = 1$ and $T_{NN} = 0$
amplitudes (dashed curve), in units of fm, with the empirical SP05
amplitudes (solid curve) \cite{Arndt} as a function of the NN
centre-of-mass angle $\theta$, in degrees, at a laboratory kinetic
energy 125~MeV.}
\end{figure}

\begin{figure}
\includegraphics[scale=1.2]{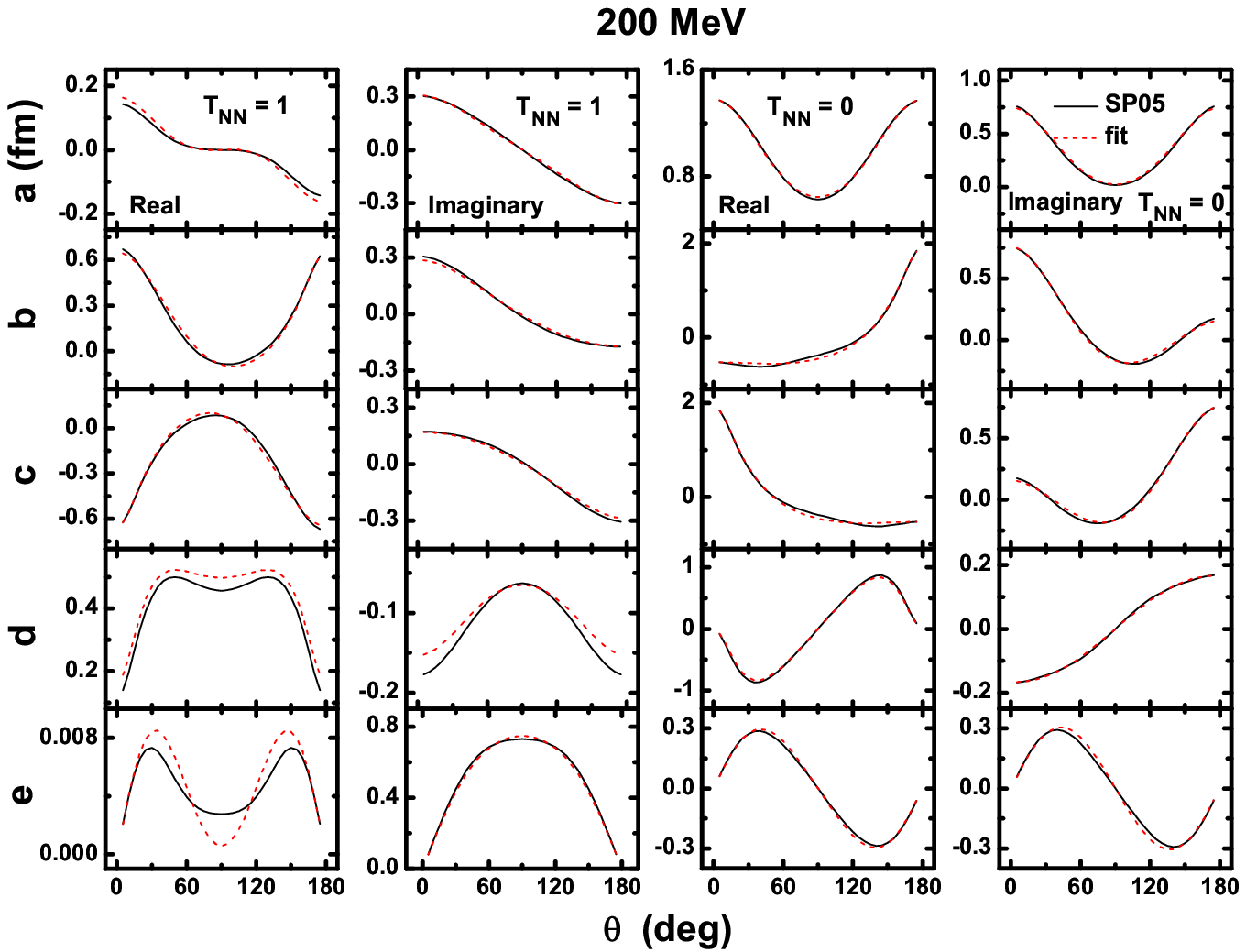} 
\caption{\label{fig:ampl-200}(Color online) Comparison of the fitted
real and imaginary ($a,b,c,d,e$) $T_{NN} = 1$ and $T_{NN} = 0$
amplitudes (dashed curve), in units of fm, with the empirical SP05
amplitudes (solid curve) \cite{Arndt} as a function of the NN
centre-of-mass angle $\theta$, in degrees, at a laboratory kinetic
energy 200~MeV.}
\end{figure}

To demonstrate the quality of the fits we compare, in
Figs.~(\ref{fig:ampl-50}) to (\ref{fig:ampl-200}), the fitted
amplitudes $a$, $b$, $c$, $d$, and $e$, for both $T_{NN} = (0, 1)$,
to the corresponding SP05 empirical values at selected energies of
50, 125 and 200~MeV. Our HLF parameter set provides a satisfactory
description of the empirical amplitudes at the latter energies.
Although not displayed, this is also the case at all other energies
between 50 and 200~MeV.

The quality of our fit is also judged by comparing $pp$ and $pn$ HLF-based and empirical scattering observables
derived from the $a,b,c,d,e$ amplitudes. In order to calculate the relevant $pp$ and $pn$ observables it is necessary
to extract the corresponding $pp$ and $pn$ amplitudes via the following relation:
\begin{eqnarray}
F_{i}(pp)\ =\ F_{i}(T = 1)
\label{e-fpp}
\end{eqnarray}
\begin{eqnarray}
F_{i}(pn)\ =\ \frac{1}{2}[F_{i}(T = 1)\ +\ F_{i}(T = 0)]\;,
\label{e-fpn}
\end{eqnarray}
and then convert them to $a,b,c,d,e$ amplitudes via
Eq.~(\ref{e-abcdesvpat}). For eventual comparison to experimental
observables, one must add Coulomb corrections (which dominate at
forward and backward scattering angles) to the $a,b,c,d,e$
amplitudes for $T_{NN}=1$: the Coulomb amplitudes were obtained by
subtracting the SAID $T_{NN}=1$ amplitudes from the SAID $pp$
amplitudes. The NN scattering observables of interest are defined as
follows \cite{Ho85}:

\begin{eqnarray}
  \sigma & = & \frac{1}{2}(|a|^2+|b|^2+|c|^2+|d|^2+|e|^2)\nonumber\\
  {\rm{P}} & = & \text{Re}(a^* e)/\sigma\nonumber\\
  {\rm{D}}& = &D_{nn}=\frac{1}{2}(|a|^2+|b|^2-|c|^2-|d|^2+|e|^2)/\sigma\nonumber\\
  {\rm{A}}_{yy} & = & \frac{1}{2}(|a|^2-|b|^2-|c|^2+|d|^2+|e|^2)/\sigma\nonumber\\
  {\rm{R}}& = &D_{s^\prime s}= \left[\text{Re}(a^*b)\cos(\alpha+\frac{\theta}{2})
                     +\text{Re}(c^*d)\cos(\alpha-\frac{\theta}{2})
                     -\text{Im}(b^*e)\sin(\alpha+\frac{\theta}{2})\right]/\sigma\nonumber\\
  {\rm{A}}& = &{\rm{D}}_{s^\prime \ell} = \left[-\text{Re}(a^*b)\sin(\alpha+\frac{\theta}{2})
                     +\text{Re}(c^*d)\sin(\alpha-\frac{\theta}{2})
                     -\text{Im}(b^*e)\cos(\alpha+\frac{\theta}{2})\right]/\nonumber\sigma
\end{eqnarray}
where
\begin{equation}
\alpha = \frac{\theta}{2}-\theta_{\rm{lab}}\;
\end{equation}
with $\theta_{\rm{lab}}$ being the laboratory scattering angle,
and $\sigma = d\sigma/d\Omega$ represents the unpolarized differential cross section.

Before comparing empirical NN scattering observables to our optimal HLF predictions,
we briefly comment on the parameter-sensitivity of the latter. We have
established that the lower energy observables are more sensitive to
variations in the values of $g^{2}$ [calculated from  $g_{0}^{2}$, $a_{g}$
and $a_{T}$ in Eq.~(\ref{e-expfunc})]. For example, for $pp$ scattering at 50~MeV,
variations of 1~\% in the $g^{2}$ values translate to a maximum
change of 6~\% on the polarization (${\rm{P}}$), the effect being
reduced for all other spin observables, and also modify the
unpolarized differential cross section ($d\sigma/d\Omega$) by 15~\%,
where the latter corresponds to a 1~\% change in the real isoscalar
tensor coupling. On the other hand, all the $pn$ observables exhibit
less sensitivity.

\begin{figure}
\includegraphics[scale=0.6]{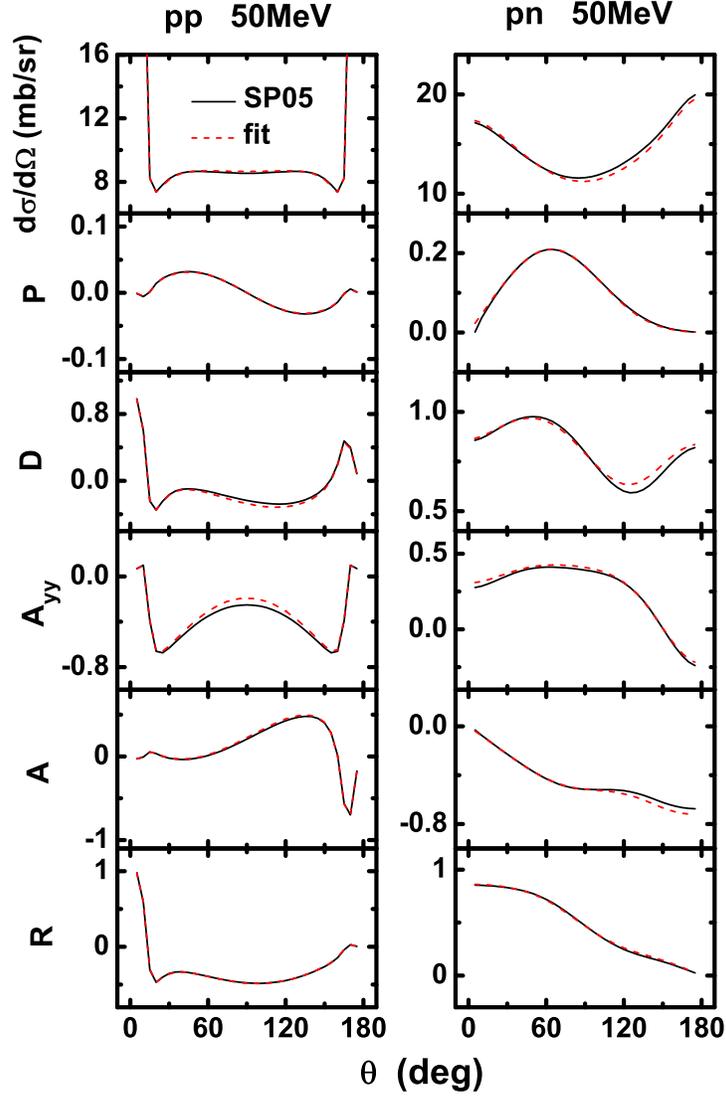} 
\caption{\label{fig:obs-50}(Color online) $pp$ and $pn$ scattering
observables versus the centre-of-mass angles $\theta$ at an incident
laboratory kinetic energy of $T_{\rm{lab}} = 50$ MeV.}
\end{figure}

\begin{figure}
\includegraphics[scale=0.6]{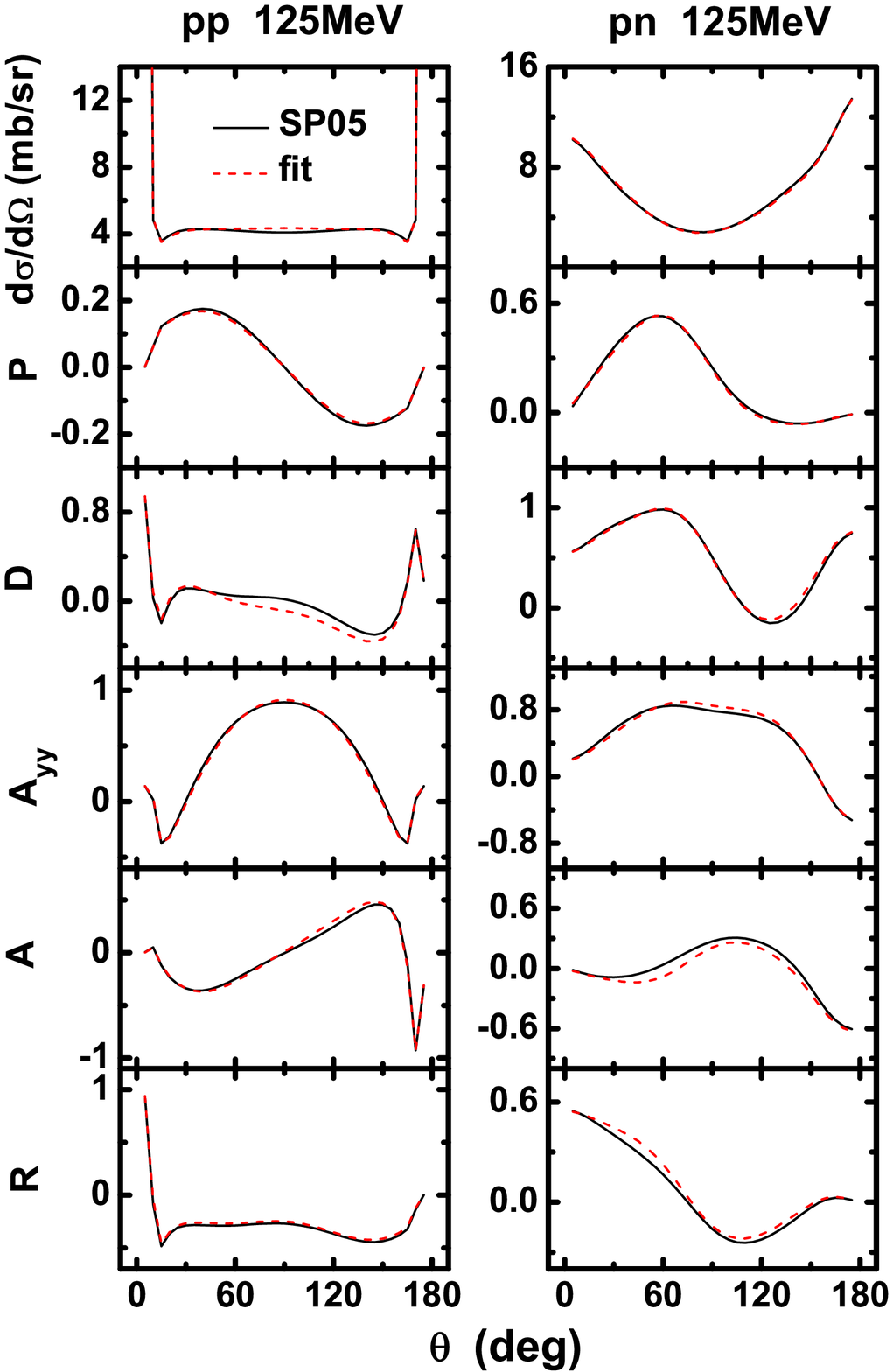} 
\caption{\label{fig:obs-125}(Color online) $pp$ and $pn$ scattering
observables versus the centre-of-mass angles $\theta$ at an incident
laboratory kinetic energy of $T_{\rm{lab}} = 125$ MeV.}
\end{figure}

\begin{figure}
\includegraphics[scale=0.6]{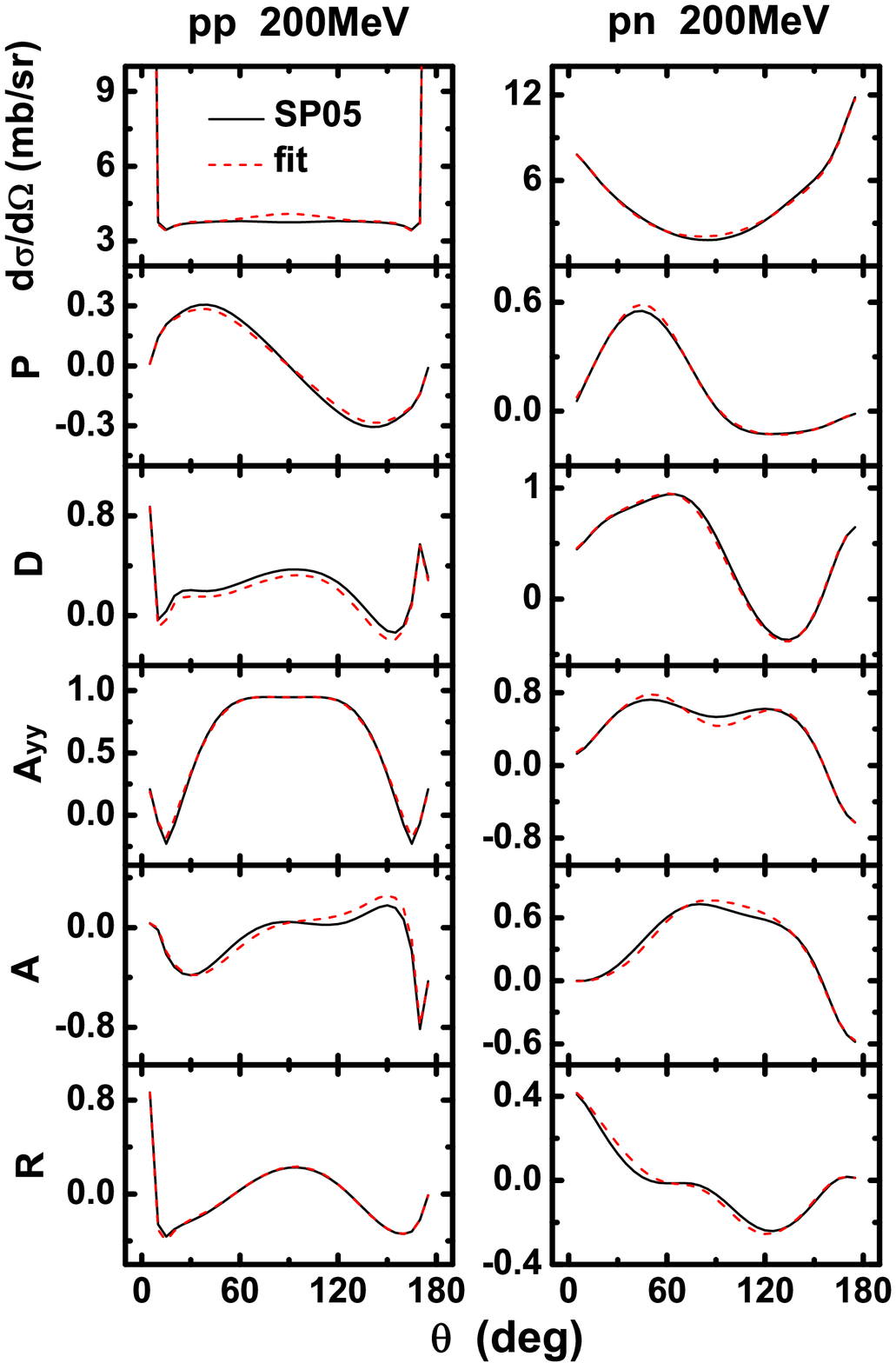} 
\caption{\label{fig:obs-200}(Color online) $pp$ and $pn$ scattering
observables versus the centre-of-mass angles $\theta$ at an incident
laboratory kinetic energy of $T_{\rm{lab}} = 200$ MeV.}
\end{figure}

Results for the unpolarized differential cross section
($d\sigma/d\Omega$), the polarization (${\rm{P}}$), the
depolarization (${\rm{D}}$), the tensor asymmetry
(${\rm{A}}_{yy}$)and triple scattering parameters ${\rm{A}}$ and
${\rm{R}}$ are shown in Fig.~(\ref{fig:obs-50}) to
(\ref{fig:obs-200}) for $pp$ and $pn$ at three different energies,
namely 50, 125 and 200 MeV. In general, our HLF parameterization
provides an excellent description of the empirical scattering
observables. In particular, the quality of our fits is just as
good,if not better, than those presented in
Refs.~\cite{Ho85,Ma96,Ma98}. Although this paper focussed on the 50
to 200~MeV range, we have also confirmed that our parametrization
provides a satisfactory description of scattering observables at a
lower enery limit of 40~MeV and a higher energy limit of 300~MeV.
The next phase of this project will be to test the predictive power
of the relativistic $t \rho$ folding procedure (for generating
microscopic relativistic scalar and vector optical potentials) for
describing elastic proton scattering from nuclei at energies lower
than 200~MeV. Indeed, at these low energies, Pauli blocking
corrections and nuclear medium modifications to the NN interaction
will play a significant role \cite{Mu87,De00,De05}. RIA predictions
as well as the importance of the latter corrections for elastic
proton scattering will be presented in a forthcoming paper.

This work is partly supported by Major State Basic Research
Developing Program 2007CB815000, the National Natural Science
Foundation of China under Grant Nos. 10435010, 10775004 and
10221003, as well as the National Research Foundation of South
Africa under Grant No. 2054166.

\end{document}